\documentstyle[12pt]{article}
\begin{document}
\renewcommand{\thepage}{ }
\begin{titlepage}
\vspace{1.5cm}
\title{
{\center Glauber dynamics in a zero magnetic field
and eigenvalue spacing statistics}
}
\author{
R. M\'elin\\
{}\\
{CRTBT-CNRS, 38042 Grenoble BP 166X c\'edex France}}
\date{}
\maketitle
\begin{abstract}
\normalsize
We discuss the eigenvalue spacing statistics of the Glauber matrix
for various models of statistical mechanics (a one dimensional Ising
model, a two dimensional Ising model, a one dimensional
model with a disordered
ground state, and a SK model with and without a ferromagnetic bias).
The dynamics of the one dimensional Ising model are integrable,
and the eigenvalue spacing statistics are non-universal.
In the other cases, the eigenvalue statistics in the high temperature
regime are intermediate between Poisson and G.O.E (with $P(0)$ of
the order of $0.5$). In the intermediate temperature regime,
the statistics are G.O.E.. In the low temperature regime,
the statistics have a peak at $s=0$.
In the low temperature regime, and for disordered systems,
the eigenvalues condense around integers, due to the fact
that the local field on any spin never vanishes.
This property is still valid for the Ising model on the Cayley tree,
even if it is not disordered.
We also study the spacing between the two largest eigenvalues
as a function of temperature. This quantity seems to be sensitive to
the existence of a broken symmetry phase.
\end{abstract}
\end{titlepage}

\newpage
\renewcommand{\thepage}{\arabic{page}}
\begin{titlepage}
\vspace{1.5cm}
\title{
{\center Dynamique de Glauber en champ magn\'etique nul et
statistique d'\'ecart entre valeurs propres}}
\maketitle
\author{
R. M\'elin\\
{}\\
{CRTBT-CNRS, 38042 Grenoble BP 166X c\'edex France}}
\date{}

\begin{abstract}
\normalsize
Nous dicutons les statistiques d'\'ecart entre valeurs propres
de la matrice de Glauber pour diff\'erents mod\`eles de m\'ecanique
statistique (mod\`ele d'Ising unidimensionnel, mod\`ele d'Ising bidimensionnel,
mod\`ele unidimensionnel avec un \'etat fondamental d\'esordonn\'e,
mod\`ele SK avec ou sans biais ferromagn\'etique).
La dynamique du mod\`ele d'Ising unidimensionnel est int\'egrable,
et la statistique d'\'ecart entre valeurs propres est non universelle.
Dans les autres cas, la statistique de valeurs propres dans le
r\'egime haute temperature est interm\'ediaire entre la loi
de Poisson et la loi G.O.E. (avec $P(0)$ de l'ordre de $0.5$).
Dans le r\'egime interm\'ediaire, les statistiques sont G.O.E..
Dans le r\'egime de basses temp\'eratures, les statistiques
pr\'esentent un pic pour $s=0$.
A basse temp\'erature et pour des syst\`emes d\'esordonn\'es,
les valeurs propres condensent autour des entiers, \`a cause
du fait que le champ local sur aucun spin ne s'annule jamais.
Cette propri\'et\'e est encore vraie pour le mod\`ele d'Ising sur l'arbre
de Cayley bien qu'il ne soit pas d\'esordonn\'e.
Nous \'etudions \'egalement l'\'ecart entre les deux
plus grandes valeurs propres en fonction de la temp\'erature.
Cette quantit\'e semble \^{e}tre sensible \`a l'existence d'une
phase avec brisure de sym\'etrie.
\end{abstract}

\end{titlepage}

\newpage
\renewcommand{\thepage}{\arabic{page}}
\setcounter{page}{1}
\baselineskip=17pt plus 0.2pt minus 0.1pt
\section{Introduction}
The ideas of level spacing statistics emerged for the first
time in the context of nuclear physics \cite{ref16}
\cite{ref17} \cite{ref18}, where Wigner proposed computing
statistical quantities from consideration of deterministic spectra.
Later, these ideas were applied to quantum systems whose
classical analogs are chaotic \cite{ref19} \cite{ref20}. The idea
of level spacing statistics is to calculate the difference
between two consecutive levels, and to study the probability
of occurrence $P(s)$ of a level spacing $s$. The different generic
behaviors of $P(s)$ are classified according to random matrix
theory \cite{ref19} \cite{ref2}.
A generic case is the integrable spectrum.
Each level is labeled by a set of quantum numbers, the energy
levels are decorrelated and the statistics are Poissonian: $P(s) =
\exp{(-s)}$.
If the number of conserved quantities is too small,
it is not possible to find a set of quantum numbers for each level,
and the levels are correlated, that is, there exists level
repulsion. The repulsion is linear and the level spacing statistics
have the Gaussian Orthogonal Ensemble (G.O.E.) shape:
\begin{equation}
P(s) = \frac{\pi}{2} s e^{- \frac{\pi}{4} s^{2}}
\label{Poisson}
{}.
\end{equation}
If time reversal invariance is broken, and if the system is chaotic,
the level spacing statistics has a Gaussian Unitary Ensemble (G.U.E.)
shape:
\begin{equation}
P(s) = \frac{32}{\pi^{2}} s^{2} e^{- \frac{4}{\pi} s^{2}}
,
\end{equation}
where the repulsion is quadratic.
The ideas of quantum chaos have been applied to various fields
of condensed matter physics, such as disordered systems \cite{ref21}.
Another field of application is strongly correlated electron
systems \cite{ref22} \cite{ref23}, where the hope is to extract more
information from finite size systems. In the present paper we wish
to analyze the dynamics of classical spin systems using
eigenvalue spacing statistics. We use here the term `eigenvalue'
rather than `level' since there are no energy levels as in quantum mechanics.
We consider the $2^{N} \times 2^{N}$ Glauber matrix, with $N$
the number of Ising spins, and diagonalize it for small clusters.
The only symmetries are the lattice
symmetries, that we treat using group theory, and the global $Z_2$ symmetry.
We can thus only study
the dynamics for a small number of sites, typically on the
order of 10 sites.
The eigenstates are not physical, except for the Boltzmann
distribution which corresponds to the upper eigenvalue $\lambda=0$.
The other eigenvectors are not probability distributions,
since the sum of their components is zero, so that their
interpretation is not obvious.

We have studied two different quantities.
The first quantity is the distance $\Delta_N(\beta)$ between the two largest
eigenvalues (one of them being zero for all $\beta$) for $N$ sites.
In the infinite temperature limit, $\Delta_N(0)=1$
(see below for a proof of this fact). Notice that in what follows
one unit of time corresponds to a single spin flip step, whereas a
Monte Carlo Step would correspond to $N$ single spin flip steps.
As a consequence, with the normalized time units the spectrum
lies in the interval $\left[ -1,1 \right]$, and the normalized
distance between the two largest eigenvalues is
$\tilde{\Delta}_N(\beta) = \Delta_N(\beta)/N$.
In particular, $\tilde{\Delta}_N(0)=1/N$ in the infinite temperature
limit.
One would expect the function $\Delta_N(\beta)$ to decrease as the inverse
temperature increases since the relaxation times are expected to be
larger for smaller temperatures.
In the presence of a broken symmetry (for instance in the case of the
two dimensional Ising model), one expects that, in the thermodynamic
limit, $\Delta_{\infty}(\beta)=0$ if $\beta>\beta_c$ since the broken symmetry
states becomes degenerate with the Boltzmann distribution in the
range of temperatures $\beta>\beta_c$.
By contrast, in the absence of a broken symmetry (for instance for a one
dimensional Ising model or for a model with a disordered ground
state), the quantity
$\Delta_{\infty}(\beta)$ should be finite,
except in the limit $\beta= + \infty$.
However, our finite size study is far from the thermodynamic limit
since we could only diagonalize the Glauber matrix for about ten sites.
We were not able to distinguish between the different conjectured behaviors
in the thermodynamic limit. Nonetheless, we study finite size effects
and show that $\Delta_N(\beta)$ decreases as the number of sites $N$
increases. For some models (one dimensional Ising model and the frustrated
one dimensional Ising model), we
find that $\Delta_N(\beta)$ is close to an exponential. In the
case of the two dimensional Ising model, $\Delta_N(\beta)$ is clearly
not exponential.

Another quantity of interest is the eigenvalue spacing
statistics of the full spectrum.
We first consider the one dimensional Ising model. In this
case, the dynamics are shown to be integrable. The corresponding eigenvalue
spacing statistics are found to be non-universal, with a peak at
$s=0$ which increases as the temperature decreases.
In the case of the two dimensional Ising model with nearest neighbor
coupling, the statistics are intermediate between Poisson and G.O.E.
for very small $\beta$, with $P(0) \simeq 0.5$. As $\beta$ increases,
the statistics evolves towards a G.O.E. shape ($\beta \sim 1$).
At low temperatures no eigenvalue repulsion exists, and the statistics
exhibit a peak at $s=0$. The weight of the peak increases with
$\beta$.
Next, we consider a frustrated one dimensional model with an extensive
entropy at low temperatures. The evolution of the eigenvalue
spacing statistics is similar to the case of the two dimensional Ising model.
In the case of the Sherrington-Kirkpatrick (SK) model,
we also have the following evolution
of the eigenvalues spacing statistics: no repulsion at very low temperatures,
repulsion for intermediate temperature ($\beta \sim 1$),
and no repulsion at low temperatures.
An important property of disordered models is that their eigenvalues
condense around integers at low temperatures. This is due to the fact
that, except for disorder realizations with zero probability measure,
the local field on any site is never zero.
However, such a behavior also exists for non random systems,
for instance for the nearest neighbor Ising model on the Cayley
tree.
\section{The Glauber matrix}
Glauber dynamics \cite{ref1} are a single spin flip dynamics
with a continuous time. If $p(\{\sigma\},t)$ is the probability
to find the spin sytem in the configuration $\{\sigma\}$
at time $t$, the master equation for the single spin flip
dynamics is
\begin{equation}
\frac{d}{dt} p(\{\sigma\},t) = -
\left( \sum_{i=1}^{N}
w_i(\{\sigma\}) \right) p(\{\sigma\},t)
+ \sum_{i=1}^{N}
w_i(\{\sigma_1,...,-\sigma_i,...,\sigma_N\})
p(\{\sigma_1,...,-\sigma_i,...,\sigma_N\},t)
{}.
\label{eq1}
\end{equation}
The single spin flip transition probabilities are defined as
the probabilities that the spin $\sigma_i$ flips from
$\sigma_i$ to $- \sigma_i$ while the other spins remain fixed.
Since the Boltzmann distribution is a fixed point
of the dynamics (\ref{eq1}), the transition probabilities
have the form
\begin{equation}
w_i(\{\sigma\}) = \frac{1}{2}(1 - \sigma_i \tanh{(\beta J
\sum_{j \in V(i)} \sigma_j))}
\label{eq10}
,
\end{equation}
where $V(i)$ is the set of neighbors of the site $i$.
Denoting the $2^{N}$ vector of the $p(\{\sigma\},t)$ as ${\bf p}(t)$,
equation (\ref{eq1}) can be written as
\begin{equation}
\frac{d}{dt} {\bf p}(t) = {\bf G}{\bf p}(t)
\label{eq2}
,
\end{equation}
where the matrix ${\bf G}$ is the Glauber matrix.
Since the Boltzmann distribution is a steady state of the dynamics, its
corresponding eigenvalue is zero regarless of temperature.
The matrix ${\bf G}$ is not symmetric. It can however be
related to a symmetric matrix ${\bf M}$. To do
so, we notice that the Glauber matrix satisfies the detailed
balance, that is ${\bf G}_{\alpha,\beta}
{\bf p}^{(0)}_{\beta} = {\bf G}_{\beta,\alpha}
{\bf p}^{(0)}_{\alpha}$, where ${\bf p}^{(0)}$ is the Boltzmann
distribution.
As a consequence,
\begin{equation}
\left( {\bf p}_{\alpha}^{(0)} \right)^{-1/2}
 {\bf G}_{
\alpha \beta} \left(
{\bf p}_{\beta}^{(0)} \right)^{1/2}
=\left( {\bf p}_{\beta}^{(0)} \right)^{-1/2}
 {\bf G}_{
\beta \alpha} \left( {\bf p}_{\alpha}^{(0)}
\right)^{1/2}
{}.
\end{equation}
We call ${\bf M}$ the matrix defined by
\begin{equation}
{\bf M}_{\alpha \beta} = \left(
{\bf p}_{\alpha}^{(0)} \right)^{-1/2}
{\bf G}_{\alpha \beta}
\left({\bf p}_{\beta}^{(0)} \right)^{1/2}
{}.
\end{equation}
Then, ${\bf M}$ is symmetric. If ${\bf p}$ is a right eigenvector
of the Glauber matrix, then
\begin{equation}
\sum_{\beta} {\bf G}_{\alpha \beta}
{\bf p}_{\beta} = \lambda {\bf p}_{\beta}
\end{equation}
is equivalent to
\begin{equation}
\sum_{\beta} {\bf M}_{\alpha \beta} \left( {\bf p}_{\beta}^{(0)}
\right)^{-1/2} {\bf p}_{\beta} = \lambda \left(
{\bf p}^{(0)}_{\alpha} \right)^{-1/2} {\bf p}_{\alpha}
,
\end{equation}
so that $\left({\bf p}_{\alpha}^{(0)} \right)^{-1/2}
{\bf p}_{\alpha}$ is an eigenvector of ${\bf M}$. We conclude
that ${\bf G}$ is diagonalizable, and that all its  eigenvalues
are real.

The spectrum in the infinite temperature
limit can be understood as
follows. If we call
\begin{equation}
| \psi \rangle = \sum_{\{\sigma\}}
f(\{\sigma\})
|\sigma_1 \rangle \otimes ... \otimes
|\sigma_N \rangle
,
\end{equation}
then the dynamics reads
\begin{equation}
\frac{d}{dt} |\psi \rangle = -
\frac{N}{2} |\psi \rangle
+ \frac{1}{2} \sum_{i=1}^{N}
\sigma_i^{x}|\psi \rangle
,
\end{equation}
so that the eigenvalues of the
Glauber matrix at infinite temperature
are of the form
\begin{equation}
\lambda = -\frac{N}{2} +
\frac{1}{2} \sum_{i=1}^{N} \mu_i
\label{eq8}
,
\end{equation}
where $\mu_i = \pm 1$.
The spectrum in the infinite
temperature limit is thus
made of eigenvalues at integer values
between $-N$ and $0$, with a
degeneracy given by the binomial
coefficients.

Another property of ${\bf G}$ is that
for bipartite lattices,
such as the square lattice or the Cayley
tree, the spectrum of ${\bf G}$
is symmetric: if $\lambda$ belongs to the
spectrum then $-N-\lambda$
is an eigenvalue also. The proof is as follows.
Let ${\bf X}\{\sigma\}$
be an eigenvector of ${\bf M}$, with an
eigenvalue $\lambda$:
\begin{equation}
\lambda {\bf X}\{ \sigma \} =
- \sum_{i=1}^{N} \frac{1}{2}(1 -
\sigma_i \tanh{(\beta J h_i)})
{\bf X}\{\sigma\} + \sum_{i=1}^{N}
\frac{1}{2 \cosh{\beta J h_i}}
{\bf X}\{\sigma_1, ..., -
\sigma_i,...,\sigma_N\}
,
\end{equation}
where $h_i$ is defined by
\begin{equation}
h_i = \sum_{j \in V(i)} \sigma_j
{}.
\end{equation}
Let ${\bf Y}\{ \sigma \}$ be defined as
\begin{equation}
{\bf Y}\{\sigma\} = (-1)^{\nu\{\sigma\}}
{\bf X}\{\tilde{\sigma}\}
,
\end{equation}
where $\nu\{\sigma\}$ is the number of up
spins in the configuration
$\{\sigma\}$.
$\{\tilde{\sigma}\}$ is deduced from
$\{\sigma\}$ by flipping the spins
of one of the two sublattices. Then,
\begin{eqnarray}
({\bf M}{\bf Y})\{\sigma\} &=&
- \sum_{i=1}^{N} \frac{1}{2}\left( 1-\sigma_i
\tanh{(\beta H h_i)} \right)
(-1)^{\nu\{\sigma\}} {\bf X}\{\tilde{\sigma}\}\\
\nonumber
&&+ \sum_{i=1}^{N} (-1)^{ \nu
\{ \sigma_1,...,-\sigma_i,...,\sigma_N \} }
\frac{1}
{2 \cosh{ (\beta J h_i) }} {\bf X}
\{\tilde{\sigma}_1,...,
-\tilde{\sigma}_i,...,\tilde{\sigma}_M\}\\
&=& (-1)^{\nu\{\sigma\}} \left[
- \sum_{i=1}^{N} \frac{1}{2}
\left(1+\tilde{\sigma}_i
\tanh{(\beta J h_i)} \right)
{\bf X}\{\tilde{\sigma}\} \right.\\
\nonumber
&&- \left. \sum_{i=1}^{N} \frac{1}{2 \cosh{(\beta J h_i)}}
{\bf X}\{\tilde{\sigma}_1,...,-
\tilde{\sigma}_i,...,\tilde{\sigma}_N\}
\right]\\
&=&-(N+\lambda) (-1)^{\nu\{\sigma\}}
{\bf X}\{\tilde{\sigma}\}
= -(N+\lambda) {\bf Y}\{\sigma\}
{}.
\end{eqnarray}
Given an eigenvector ${\bf X}$ for the
eigenvalue $\lambda$,
we have built an eigenvector ${\bf Y}$
for the eigenvalue $-N-\lambda$.

The difference between (\ref{eq2}) and
the Schr\"odinger equation is that
quantum mechanics preserves the scalar
product, leading to Hermitian
Hamiltonians. More over, the space of
physical states is a Hilbert space,
and each state of the Hilbert state is
physical. In the case of the Glauber
matrix, no vector space is present
in the sense that the sum of
two probability distributions is not a
probability distribution.
However, some quantities are
conserved by the dynamics. It is easy
to show that the eigenvectors
of ${\bf G}$ for the non-zero eigenvalues
have the property that
\begin{equation}
\sum_{\{\sigma\}} {\bf p}\{\sigma\} = 0
{}.
\end{equation}
This is a simple consequence of the fact that
the Glauber matrix preserves the quantity
\begin{equation}
\sum_{\{\sigma\}} {\bf p}\{\sigma\}
{}.
\end{equation}
\section{One dimensional Ising model}
\subsection{Integrability of the dynamics}
In the case of the one dimensional model, the Glauber dynamics is
integrable. To show this, we follow Glauber and write evolution
equations for the correlation functions.
We call $R^{(n)}_{i_1,...,i_n}(t)$ the $n$-point correlation function,
with $i_{\alpha} \ne i_{\beta}$ if $\alpha \ne \beta$,
that is
\begin{equation}
R^{(n)}_{i_1,...,i_n}(t) = \langle \prod_{\alpha}
\sigma_{i_{\alpha}}(t) \rangle
{}.
\end{equation}
Then following Glauber, we write the evolution equation of
$R^{(n)}_{i_1,...,i_n}(t)$ under the form
\begin{equation}
\frac{d}{dt} R^{(n)}_{i_1,...,i_n}(t) = - 2 \langle
\sigma_{i_1}(t) ... \sigma_{i_n}(t) \left(
w_{i_1}(\{\sigma\})+...+w_{i_n}(\{\sigma\}) \right) \rangle
\label{eq4}
,
\end{equation}
where the transition probabilities are given by (\ref{eq10}).
We notice that in the one dimensional case, each spin
has two neighbors, so that $w_i(\{\sigma\})$ can be written as
\begin{equation}
w_i(\{\sigma\}) = \frac{1}{2} \left( 1 - \frac{\gamma}{2}
\sigma_i(\sigma_{i+1}+\sigma_{i-1}) \right)
\label{eq3}
,
\end{equation}
with $\gamma = \tanh{2 \beta J}$.
Here, we take periodic boundary conditions, but the case of open boundary
conditions is similar.
Inserting (\ref{eq3}) into (\ref{eq4}) we get
\begin{equation}
\frac{d}{dt} R^{(n)}_{i_1,...,i_n}(t)
= -n  R^{(n)}_{i_1,...,i_n}(t) + \frac{\gamma}{2}
\sum_{\epsilon=\pm 1} \sum_{\alpha=1}^{n}
\langle \sigma_{i_{\alpha}+ \epsilon} \prod_{\beta \ne \alpha}
\sigma_{i_{\beta}} \rangle
{}.
\label{eq5}
\end{equation}
The terms with $\epsilon=1$ collect the right neighbors,
and $\epsilon=-1$ corresponds to the left neighbors.
The correlation function in (\ref{eq5}) leads to
a (n-2)-point correlator if $\exists \beta, i_{\alpha +
\epsilon} = i_{\beta}$  or to a $n$-point correlator if not.
The expression (\ref{eq5}) can be brought under the form
\begin{eqnarray}
\label{eq6}
\frac{d}{dt} R^{(n)}_{i_1,...,i_n}(t)
&=& -n  R^{(n)}_{i_1,...,i_n}(t) + \frac{\gamma}{2}
\sum_{\epsilon= \pm 1} \sum_{\alpha=1}^{n} \left(
\sum_{\beta=1}^{n} \delta_{i_{\beta},i_{\alpha}+\epsilon}
R^{(n-2)}_{i_1,...,i_{\alpha-1},i_{\alpha+1},...,i_{\beta-1},
i_{\beta+1},...,i_n}(t) \right.\\
&&+ \left. \left( 1 - \sum_{\beta=1}^{n} \delta_{i_{\beta},
i_{\alpha}+\epsilon} \right)
R^{(n)}_{i_1,...,i_{\alpha-1},i_{\alpha}+\epsilon,
i_{\alpha+1},...,i_n}(t) \right)
\nonumber
{}.
\end{eqnarray}
If none of the sites $i_1,...,i_n$ are neighbors,
the term containing
$R^{(n-2)}$ vanishes. However, if at least two sites
in the set $i_1,...,i_n$ are neighbours,
we have to take into account a term
containing $R^{(n-2)}$ in the evolution of $R^{(n)}$.
It is clear that (\ref{eq6}) is nothing but a rewritting of
(\ref{eq1}) in the case where all the sites have only two
neighbors. The number of distinct correlation functions is
\begin{equation}
\sum_{n=0}^{N} {N \choose n} = 2^{N}
,
\end{equation}
which is equal to the number of spin configurations.
The system (\ref{eq6}) is integrable. Glauber gives the explicit
solution for $R^{(1)}_i(t)$.
The equation giving $d R^{(2)}(t)/dt$ contains only linear combinations
of $R^{(2)}$.
In order to solve for the three points
correlation functions, we inject Glauber's solution into the
evolution equation for $R^{(3)}$, diagonalize the associated
matrix and get a first order differential equation,
which is explicitly integrable and yields
the second order correlation functions. The entire
hierarchy can be solved by this method since $ d R^{(k)}/dt$ does
not contain $R^{(p)}$ with $p>k$.

\subsection{Eigenvalue spacing statistics}
\label{this section}
In order to calculate the eigenvalue spacing statistics, we need to
take all the symmetries of the lattice into account. Here, the
symmetries are so obvious that we do not require a group
theory treatment. We work with an open Ising chain. This graph
is invariant under the reflection and the identity operators.
We denote  the basis of our `Hilbert' space as $\{\sigma\}\rangle$.
We use quotes since there is no vector space structure on the
probability distributions. However, to diagonalize the
Glauber matrix, we can use an analogy to quantum mechanics.
If $R$ is the reflection operator, we form the combinations
\begin{equation}
| \{ \sigma \} \rangle_{\epsilon} =
\frac{1}{\sqrt{2}} \left( |\{ \sigma \}\rangle + \epsilon
R | \{ \sigma \} \rangle \right)
{}.
\end{equation}
This operation leads to states with a well defined behavior
under the reflection. The resulting state is either symmetric
($\epsilon=1$) or antisymmetric ($\epsilon=-1$).
The antisymmetric state may be zero if
$|\{\sigma\} \rangle$ is invariant under the reflection.
The dimension of the antisymmetric sector is
\begin{equation}
\frac{1}{2} \left( 2^{N} - 2^{\left[ \frac{N+1}{2} \right]} \right)
,
\end{equation}
and the dimension of the symmetric sector is
\begin{equation}
\frac{1}{2} \left( 2^{N} + 2^{\left[ \frac{N+1}{2} \right]} \right)
,
\end{equation}
where $\left[ \right]$ denotes the integer part.
Finally, we take into account the global $Z_2$ symmetry of the
Glauber matrix. [The matrix elements of the Glauber matrix are invariant
under the transformation $\{\sigma_i\} \rightarrow \{-\sigma_i\}$].
Taking into account all the symmetries, we diagonalize
the Glauber matrix in the reflection symmetric and the antisymmetric sectors,
and in the $Z_2$ symmetric and antusymmetric sectors.
The evolution of the eigenvalues of one sector as a function of the
inverse temperature is plotted in figure \ref{Fig1} (a).
No avoided crossings are present, which is what is expected for an
integrable system.
The difference between the two largest eigenvalues is plotted
on figure \ref{Fig2} as a function of the inverse temperature.
Eventhough some deviations are visible, the behavior of
$\Delta_N(\beta)$ is close to an exponential decay. Finite size effects
are visible: $\Delta_N(\beta)$ decreases as the number of sites $N$
increases.

The eigenvalue spacing statistics are found to be non universal.
For instance, $P(s)$ is plotted on figure \ref{Fig3} for
$\beta=1$.The height of the peak at $s=0$ decreases as the temperature
decreases. Eventhough the dynamics is integrable, the eigenvalue
spacing statistics are not of the Poisson form (\ref{Poisson}).
Such a behavior for the spectrum of integrable systems has already been
found in the context of integrable quantum fluids (see \cite{ref23}).

\section{Bidimensional Ising model}
\subsection{Dynamics}
In this case, the dynamics are no longer integrable. The evolution
of the correlation function is not a linear equation, as it
was in the case of the Ising chain. This is essentially due to the
fact that, with four neighbors, one has to introduce a cubic term
in $w_i(\{\sigma\})$ given by equation (\ref{eq3}):
\begin{equation}
\tanh{(\beta J (\sigma_1 + \sigma_2 + \sigma_3
+ \sigma_4))} = \alpha (\sigma_1 + \sigma_2 + \sigma_3
+ \sigma_4) + \alpha' (\sigma_1 + \sigma_2 + \sigma_3
+ \sigma_4)^{3}
,
\end{equation}
with
\begin{eqnarray}
\alpha &=& \frac{1}{12} \left( 8 \tanh{2 \beta J} - \tanh{4 \beta J} \right)\\
\alpha' &=& \frac{1}{48} \left( \tanh{4 \beta J} - 2 \tanh{2 \beta J} \right)
{}.
\end{eqnarray}
In the one dimensional case, we could integrate the dynamics
because $d R^{(n)}/dt$ was only a function of $R^{(k)}$ with
$k \le n$. In the two dimensional case $d R^{(n)}/dt$ is
also a function of $R^{(k)}$ with $k>n$, so that the
hierarchy is no longer integrable by this method. It is not because
one does not know how to solve the dynamics that these dynamics are
not integrable. The analysis of the spectral statistics of the
Glauber matrix may be useful to determine whether or not
there exists some conserved quantities in the Glauber dynamics
of the the Ising model.

\subsection{Use of group theory}
We use group theory to find the symmetries of the
clusters for which we shall diagonalize the Glauber matrix.
Notice that we are restricted to small sizes since the size of
the `Hilbert' space is equal to $2^{N}$. In practice, and to
have reasonable execution times, we are restricted to $N \le 13$.
The first step is to determine the symmetry group of the
lattice, that is to enumerate all the permutations that
leave the lattice invariant. To do so, we do not test all the
$N!$ possible permutations since the computation time may be huge.
Instead, we use the following procedure.
We first label the lattice sites and give the list of bonds.
We determine all the possible images $\sigma(1)$ of the site $1$, that is
all the $N-1$ sites. Then, for each of the possible images of
the site $1$, we determine the images $\sigma(2)$
of site $2$ which respect the lattice symmetry: if there is a bond
between $1$ and $2$, there must be a bond between $\sigma(1)$
and $\sigma(2)$. If there is no bond between $1$ and $2$, there must
be no bond between $\sigma(1)$ and $\sigma(2)$. At this point,
we have a list of potential permutations beginning with
$\sigma(1)$ and $\sigma(2)$. Then, we determine all the possible
images of site $3$ which leave the lattice invariant.
We thus get a tree structure, but, during the construction,
some branches shall stop. At the end of the process, that is
when $\sigma(N)$ has been determined, we get all the permutations
which leave the graph invariant.
The second step is to determine the classes and the table of
characters of the group. We use the program in \cite{ref3},
which automatically determines the classes and the table
of characters.
In a third step, we have to determine the size of the
blocs corresponding to the irreducible representations, and
how many times a given irreducible representation appears.
The dimension of the blocs corresponding to the
irreducible representation $(j)$ is equal to
\begin{equation}
Dim^{(j)} = \frac{1}{h} \sum_{g \in G} \chi(\tilde{g})
\chi^{(j)}(\tilde{g})
\label{eq7}
{}.
\end{equation}
$\tilde{g}$ is the representation of the group element $g$
in the `Hilbert' space, $h$ is the cardinal of the group $G$,
$\chi(\tilde{g})$ is the trace of $\tilde{g}$ and
$\chi^{(j)}(g)$ is read from the table of characters at the intersection
of the line corresponding to the representation $(j)$ and the column
of the class of $\tilde{g}$.
The fourth step is to implement a Gram-Schmidt procedure to determine
the basis of one block corresponding to the representation $(j)$.
We first use the projector
\begin{equation}
P^{(j)} = \sum_{g \in G} \chi^{(j)}(\tilde{g}) \tilde{g}
{}.
\end{equation}
A basis element of the Hilbert space is coded as a binary number of size $N$.
Zero corresponds to a down spin, and 1 codes an up spin. In order
to label the basis vectors, we use the decimal representation
of the binary number of size $N$. We denote the corresponding
vector $|\psi_k \rangle$.
The procedure consists of scanning all the
states $|\psi_k \rangle$ and to determine $k_0$ such as
$P^{(j)} |\psi_k \rangle =0$ if $k<k_0$ and $P^{(j)}|\psi_{k_0} \rangle
\ne 0$. The state $P^{(j)} |\psi_{k_0} \rangle$ is the first
vector of the basis that we are looking for. Once we have found the
first vector of the basis, we continue to scan all the states
$|\psi_k \rangle$, but we project them with
\begin{equation}
P^{(j)}_0 = \sum_{g \in G} \langle \psi_{k_0}| \tilde{g} |
\psi_{k_0} \rangle \tilde{g}
{}.
\end{equation}
If $P^{(j)}_0 |\psi_k \rangle = 0$, we forget about
$|\psi_k \rangle$ and project $|\psi_{k+1} \rangle$.
If $P^{(j)}_0 |\psi_k \rangle \ne 0$, we try to incorporate
$P^{(j)}_0 |\psi_k \rangle$ into the basis using a modified
Gram-Schmidt procedure \cite{ref5}.
If $P^{(j)}_0 |\psi_k \rangle$ is a linear combination
of the basis vectors, then we discard it and
project $|\psi_{k+1} \rangle$. If it is not, we incorporate
it into the basis, after having orthogonalized it, and
we make the projection test for $|\psi_{k+1} \rangle$. At the
end of the procedure, the dimension of the basis must be
equal to (\ref{eq7}). We note that it is not possible to
store all the components of the orthonormal basis because
of limited storage capacity. In order to
save memory, we stored only the non zero components.
The fifth step is to take into account the global $Z_2$ symmetry.
The sixth and last step is to diagonalize the Glauber matrix using
the basis that has been determined at the fourth step.
The size of the matrices to be diagonalized are small enough, so
that we can use the Jacobi method.

\subsection{Results}
We work with a 3x4 lattice with periodic boundary conditions.
The number of representations is equal
to 15, and the maximal block dimension is 335.
The spectrum in a given sector of symmetry of the 3x4 square
lattice is pictured in Figure \ref{Fig1} (b) as a function
of the inverse temperature. In the limit
$\beta \rightarrow 0$, we recover degeneracies for integer
eigenvalues (see relation (\ref{eq8})). As the inverse temperature
increases, the degeneracies present for $\beta=0$ are lifted,
but the eigenvalues from two different degeneracies are not free
to cross, due to eigenvalue repulsion.

We studied the evolution of $\Delta_N(\beta)$ as a function of
$\beta$ for the 3x4 lattice and for the 3x3 lattice.
The result is plotted in figure \ref{Fig5}. Finite size effects are
visible; $\Delta_N(\beta)$ decreases if $N$ increases,
and the evolution of $\Delta_N(\beta)$ is clearly non exponential.
In the thermodynamic limit,
$\Delta_{\infty}(\beta)=0$ if $\beta>\beta_c$ and
$\Delta_{\infty}(\beta)>0$ if $\beta<\beta_c$, which means that
$\ln{\Delta_{\infty}(\beta)} \rightarrow - \infty$ if
$\beta \rightarrow \beta_c^{-}$.
Our finite size study is consistent with such a behavior. However,
it would be useful to analyze larger samples, which will be done
in the near future.

We now discuss the shape of the eigenvalue spacing statistics $P(s)$.
If the inverse temperature is very small, the degeneracies of the
$\beta=0$ case are lifted and the once degenerate eigenvalues
spread out linearly.
The corresponding statistics are plotted
in Figure \ref{Fig6} for $\beta=0.01$. In this case, we find $P(0)
\simeq 0.5$ and the statistics are close to the Poisson law for large $s$.
If $\beta$ increases, one reaches the eigenvalue repulsion regime (see figure
\ref{Fig1} (b)). Linear eigenvalue
repulsion is visible on figure \ref{Fig6} since,
for $\beta=1$, $P(s) \sim s$ for small $s$. However, for $s$ of order unity,
large deviations to the G.O.E. law occur.
For large $\beta$, the statistics are not universal (see Figure \ref{Fig7}
for $\beta=10$), with a peak at $s=0$. The weight of this peak increases
as $\beta$ increases.

\section{Frustrated one dimensional model}
\subsection{The model}
We consider the one dimensional antiferromagnetic Ising model with
antiferromagnetic next-nearest-neighbor interactions.
This model can be seen as a succession of triangles, as pictured
in Figure \ref{Fig8} and can be solved via
a transfer matrix formalism, with the sites gathered as
shown on figure \ref{Fig8}. The transfer matrix has the
form
\begin{equation}
\label{eq9}
T =
\left(
\begin{array}{clcr}
A & B\\
B & A
\end{array}
\right)
,
\end{equation}
with
\begin{equation}
A =
\left(
\begin{array}{clcr}
e^{-4 \beta J} && e^{-2 \beta J}\\
e^{2 \beta J} && 1
\end{array}
\right)
\mbox{ , }
B =
\left(
\begin{array}{clcr}
e^{2 \beta J} && 1\\
1 && e^{2 \beta J}
\end{array}
\right)
{}.
\end{equation}
In (\ref{eq9}), the states are ordered in the form $|\uparrow,
\uparrow \rangle$, $|\uparrow,\downarrow \rangle$, $|\downarrow,
\downarrow \rangle$, $|\downarrow,\uparrow \rangle$.
Because of the form (\ref{eq9}) of the transfer matrix,
if $(\psi,\varphi)$ is and eigenvector of $T$, then
$(\psi+\varphi,\psi+\varphi)$ and $(\psi-\varphi,\psi-\varphi)$
are eigenvectors of $T$ for the same eigenvalue,
so that the eigenvalues of $T$ are the eigenvalues
of $A+B$ and $A-B$, and the initial 4x4 problem
is reduced to two 2x2 problems, due to the time reversal invariance.
The partition function is simply
$Z_N = Tr T^{N/2}$ for a $N$ site chain ($N$ is assumed
to be even). The zero temperature entropy is found to be extensive,
of the form $S(0)/N = \ln{2}/2$, whereas in the corresponding ferromagnetic
problem, the entropy is finite at zero temperatures.
This one dimensional antiferromagnetic model has thus the same properties
as the triangular antiferromagnet \cite{ref6}, namely the number of
zero temperature ground states is proportional to $\exp{\alpha N}$,
with $\alpha$ a constant.

\subsection{Results}
We work with an open chain version, so that the only symmetries are
the inversion and the global $Z_2$ symmetry.
We have already explained how to treat these symmetries (section
\ref{this section}).

The evolution of $\Delta_N(\beta)$ as a function of the inverse
temperature $\beta$ is plotted in Figure \ref{Fig9} for different
values of $N$. We observe finite size effects:
$\Delta_N(\beta)$ decreases with $N$. The variations of $\Delta_N(\beta)$
as a function of $\beta$ are linear in a semi-log plot, so that
$\Delta_N(\beta)$ decreases exponentially with $\beta$.

The evolution of the eigenvalues of the Glauber matrix as a function
of the inverse temperature is plotted in figure \ref{Fig1} (c).
Level repulsion is visible for $0.2 < \beta < 1$.

The eigenvalue spacing statistics in the high temperature regime
is plotted in Figure \ref{Fig11} ($\beta=0.01$).
In this range of temperatures, the statistics are intermediate between
Poisson and G.O.E. $P(s)$ has a maximum, but is occurs for
smaller separations than in the G.O.E. case.
If the inverse temperature increases, the maximum occurs for
larger separations, and the eigenvalue spacing statistics is close
to the G.O.E. law for $\beta=1$ (see Figure \ref{Fig11}).
For larger temperature ($\beta=3$ on Figure \ref{Fig12}),
$P(0)$ is not zero (for instance, $P(0) \simeq 0.4$ for $\beta=3$)
and $P(s)$ is close to the Poisson law for large spacings.

\section{SK model}
\subsection{The model}
This model was proposed in 1975 as an `exactly solvable'
spin glass model \cite{ref7}. For general reviews on
the problems of spin glasses, we refer the reader to references
\cite{ref8} \cite{ref9} \cite{ref10}. The SK model is defined
by the disordered, infinite range interaction Hamiltonian
\begin{equation}
H = \sum_{\langle i,j \rangle} J_{ij} \sigma_i \sigma_j
,
\end{equation}
with quenched random interactions $J_{ij}$ with a gaussian
distribution
\begin{equation}
P(J_{ij}) = \left( \frac{N}{2 \pi J^{2}} \right)^{1/2}
\exp{\left(- \frac{(N J_{ij} - \delta)^{2}}{2 J^{2}}\right) }
,
\end{equation}
In this section, we study the symmetric case $\delta=0$.
The case $\delta >1$ is the object of section \ref{sec}.
We refer the reader to the reviews previously quoted for
the solution of this model. For $T<J$, the model is a spin glass,
and a paramagnet for $T>J$.
In the glass phase, the model has a large number of thermodynamic
phases, with no symmetry connecting ground states.
The different ground states are separated by large barriers,
proportional to the system size.
An other feature of glassiness is the presence of aging, associated
with slow relaxation processess.

\subsection{Eigenvalue spacing statistics}
In the case of the SK model,
we do not need to look for the lattice symmetries
since, except for some special cases of zero probability,
the random infinite range interactions break all the lattice
symmetries. The only symmetry to be taken into account is the
global $Z_2$ symmetry.

The evolution of the eigenvalues as a function of the inverse
temperature is plotted in Figure \ref{Fig13} (a) in a given symmetry sector.
We observe that, in the limit $\beta \rightarrow + \infty$, the eigenvalues
condense around integers. This property is due to the fact that, in this
limit, the symmetric representation ${\bf M}$, of ${\bf G}$, is diagonal
since the local field never vanishes (except for some disorder configurations
of zero probability measure), and that the diagonal coefficients are integers.
This cluster property in the zero temperature limit is an effect of
disorder, and is  not related to the existence of glassiness.
However, as we shall see in section \ref{tree}, we find a similar
behavior for a glassy ferromagnetic system with no disorder !

The evolution of $\overline{\Delta_N(\beta)}$ as a function of
$\beta$ is plotted in Figure \ref{Fig14}. The variations of
$\overline{\Delta_N(\beta)}$ as a function of $\beta$ are approximately
linear in a semi-log plot, so that  $\overline{\Delta_N(\beta)}$ decays
exponentially as a function of $\beta$.

The eigenvalue spacing statistics $\overline{P(s)}$ in the high temperature
regime are intermediate between the Poisson law and the G.O.E. law
(see Figure \ref{Fig15} for $\beta=0.01$). In this regime,
$\overline{P(0)} \simeq 0.5$
and $\overline{P(s)}$ is close to the Poisson law for large separations.
If the inverse temperature $\beta$ increases, the maximum of
$P(s)$ occurs for larger separations (see Figure \ref{Fig15} for
$\beta=0.1$) and finally $\overline{P(s)}$ reaches the G.O.E. law
(see figure \ref{Fig16} for $\beta=1$). In the very low temperature
limit, the statistics are close to the Poisson law, with a peak
for $s=0$ (see Figure \ref{Fig16} for $\beta=10$). The height of this
peak increases as $\beta$ increases.

\subsection{Zero temperature density of zero eigenvalues}
At zero temperature, the eigenvalues of the Glauber matrix of the SK
model are integers, and the Glauber matrix is diagonal in the natural
basis, excepted for some disorder configurations of zero probability
measure. As a matter of fact, some eigenvalues are zero. The states
corresponding to the zero temperature zero eigenvalue are the metastable
states of the SK model, and their degeneracy is
expected to increase
exponentially with the system size, as expected from the number
of solutions to the TAP equations \cite{ref8}.
It is thus interesting to calculate the density of zero eigenvalues
in the zero temperature limit as a function of the system size, even if
our approach is restricted only to small system sizes. First, we
analytically treat the case of the 3 site SK model, as a warm up exercise.

\subsubsection{Warmup: the 3 site SK model}
We call $\sigma_i$ the spin variable on site $i$ ($i=0,1,2$), and
$J$ ($K$,$L$) the random bond between $\sigma_0$ and $\sigma_1$
($\sigma_0$ and $\sigma_2$, $\sigma_1$ and $\sigma_2$).
Since the local field never vanishes, the Glauber matrix is diagonal,
and the diagonal coefficients are
\begin{eqnarray}
\Delta_1 &=&\Delta(+,+,+)
= \Delta(-,-,-) = -\frac{3}{2} + \frac{1}{2} \epsilon (J+K) +
\frac{1}{2} \epsilon (J+L) + \frac{1}{2} \epsilon (K+L)\\
\Delta_2 &=&\Delta(+,+,-)
= \Delta(-,-,+) = -\frac{3}{2} + \frac{1}{2} \epsilon (J-K) +
\frac{1}{2} \epsilon (J-L) - \frac{1}{2} \epsilon (K+L)\\
\Delta_3 &=& \Delta(+,-,+) =
\Delta(-,+,-) =-\frac{3}{2} + \frac{1}{2} \epsilon (K-J) -
\frac{1}{2} \epsilon (J+L) + \frac{1}{2} \epsilon (K-L)\\
\Delta_4 &=&
\Delta(-,+,+) =
 \Delta(+,-,-) =-\frac{3}{2} - \frac{1}{2} \epsilon (J+K) -
\frac{1}{2} \epsilon (J-L) - \frac{1}{2} \epsilon (K-L)
,
\end{eqnarray}
where $\epsilon(x)$ denotes the sign of $x$.
The gaussian bond distribution is replaced by
\begin{equation}
P_{\delta'}(J) = \theta(\frac{1}{\sqrt{3}}+\delta'-J)
\theta(J+\frac{1}{\sqrt{3}} - \delta')
\end{equation}
in order to analytically perform the forthcoming integrations.
The variable $\delta'$ is equal to $\delta/3$.
After lengthy but straightforward calculations, we get
the probability ${\cal P}_i(\delta')$ that $\Delta_i$ ($i=0,1,2,3$)
takes the value zero:
\begin{itemize}
\item if $\delta'>1/\sqrt{3}$, ${\cal P}_0(\delta')=1$ and
${\cal P}_1(\delta')=0$.
\item if $0 \le \delta' \le 1/\sqrt{3}$,
\begin{eqnarray}
{\cal P}_0(\delta') &=& \frac{1}{4} + \frac{3 \sqrt{3}}{4} \delta'
+ \frac{9}{4} \delta'^{2} - \frac{9 \sqrt{3}}{4} \delta'^{3}\\
{\cal P}_1(\delta') &=& \frac{1}{4} - \frac{\sqrt{3}}{4} \delta'
- \frac{3}{4} \delta'^{2} + \frac{3 \sqrt{3}}{4} \delta'^{3}
{}.
\end{eqnarray}
\item if $-1/\sqrt{3} \le \delta' \le 0$,
\begin{eqnarray}
{\cal P}_0(\delta') &=& \frac{3 \sqrt{3}}{4} \left( \frac{1}{\sqrt{3}}
+ \delta' \right)^{3}\\
{\cal P}_1(\delta') &=& \frac{1}{4} - \frac{\sqrt{3}}{4} \delta'
- \frac{3}{4} \delta'^{2} - \frac{\sqrt{3}}{4} \delta'^{3}
{}.
\end{eqnarray}
\item finally, if $\delta' \le - 1\sqrt{3}$, ${\cal P}_0(\delta')=
0$ and ${\cal P}_1(\delta') = 1/3$.
\end{itemize}
Moreover, ${\cal P}_1={\cal P}_2={\cal P}_3$.
We observe that ${\cal P}_0 + 3 {\cal P}_1 = 1$ regardless of the value
of the ferromagnetic bias $\delta'$. In the ferromagnetic region,
only one state (ferromagnetic) has a zero eigenvalue. In the frustrated
region, three `disordered' configurations share the zero eigenvalue,
while the probability of the ferromagnetic state is vanishing.
For this three site toy model, the probability of finding a zero eigenvalue
is a constant as a function of the ferromagnetic bias. This example
shows that two mechanisms are responsible for the appearance of
zero eigenvalues: (frustration + disorder) or broken symmetry.
Notice that both frustration and disorder are important here since
without disorder, the Glauber matrix would not be diagonal in the
zero temperature limit.

\subsection{$N$ site SK model}
We studied the total probability ${\cal P}_N(\delta)$ of finding a zero
eigenvalue at zero temperature. This quantity is plotted in Figure
\ref{Fig17} in the frustrated region ($\delta=-2$), in the
zero bias case ($\delta=0$) and in the ferromagnetic region
($\delta=2$). We observe that in both cases  ${\cal P}_N(\delta)$
increases with $N$, but more slowly in the ferromagnetic case
than in the zero bias case. This suggests that the (frustration +
disorder) mechanism to generate zero eigenvalues of the Glauber matrix at
zero temperature is more efficient than the broken symmetry mechanism,
in the limit of large $N$. Moreover, we see that the variations of
${\cal P}_N(\delta)$ as a function of $N$ are compatible with an exponential
growth. However, we are restricted to small systems, so that
it would be interesting to do the same study for larger systems.
This shall be done in the near future.
Another quantity of interest is the complete distribution
function $P(\Delta_N(\beta))$. This quantity is plotted in Figure
\ref{Fig18} in the low temperature and high temperature regime.
We observe that, in the low temperature regime,
the standard deviation of $\Delta_N(\beta)$ is of the same order
as $\Delta_N(\beta)$, whereas in the high temperature regime,
the standard deviation is small compared to the mean value.

\section{SK model with a ferromagnetic bias}
\label{sec}
We now consider the SK model with a ferromagnetic bias $\delta > 1$.
In this regime, the SK model possesses a paramagnetic/ferromagnetic
transition as a function of temperature.
The spectrum of the Glauber matrix is plotted on Figure \ref{Fig13} (b)
as a function of the inverse temperature $\beta$. Level repulsion is
visible, and the eigenvalues are attracted by integers in the limit
of large $\beta$, which is consistent with the fact that the Hamiltonian
is disordered. We plotted on Figure \ref{Fig14}
$\overline{\Delta_N(\beta)}$ for
different values of the bias $\delta$. We observe that
$\overline{\Delta_N(\beta)}$ decreases as the ferromagnetic bias increases.
This is due to the fact that the paramagnetic/ferromagnetic
transition temperature of the biased SK model increases with
the bias $\delta$ \cite{ref7}, so that the inverse transition temperature
decreases as $\delta$ increases, leading to a decrease
of $\overline{\Delta_N(\beta)}$ as $\delta$ increases.
Moreover, the variations of $\overline{\Delta_N(\beta)}$ become
non-exponential if the ferromagnetic bias increases (see the case
$\delta=3$ of Figure \ref{Fig14}), which is consistent with what has
been observed in the case of the two dimensional Ising model.
The eigenvalue spacing statistics are plotted in Figure
\ref{Fig19} and Figure \ref{Fig20}. Level repulsion appears
even for small values of $\beta$.

\section{Ferromagnetic Cayley tree}
\label{tree}
Trees were introduced in statistical physics as early
as the 1930's in order to implement mean field theories \cite{ref11}
\cite{ref12}. In this case, only the central spin was considered
and the border of the tree (which represents a finite fraction of the spins)
was sent to infinity. This limit is
the Bethe-Peierls limit. However, it is possible to develop
statistical physics on a Cayley tree, with the border included.
This was done in the seventies, and a continuous transition
was discovered \cite{ref13} \cite{ref14}. The dynamics was studied
only recently \cite{ref15}. A cross-over (not a transition like in the
SK model) to a glassy regime was then
found below the temperature scale $T_g \sim J / \ln{n}$, with $n$ the
number of generations. The temperature scale $T_g$ decreases in the
limit $n \rightarrow + \infty$, but very slowly, so that the glassy
regime exists even in the macroscopic regime.
The zero temperature barriers scale like $J \ln{n}$, with $n$ the number
of generations.
If the coordination of the bulk sites is equal to 3, the local field
of the bulk sites cannot vanish. This property is reminiscent on
the behavior of disordered systems. The analysis of the spectrum
of the Glauber matrix at low temperatures reveals that the clustering
property also exists for the Cayley tree, as shown in Figure
\ref{Fig22}, where we plotted the eigenvalues of the Glauber
matrix on the Cayley tree at low temperature. This indicates a
striking analogy between the ferromagnetic Ising model on the
Cayley tree and disordered systems.

\section{Conclusion}
Since the contents of this paper were already summarized in the
introduction, we end here with some final remarks and open questions.
The one dimensional case is very special due to the existence
of an underlying integrable dynamics. In the other cases,
we have the following scenario for the evolution of the eigenvalue
spacing statistics as a function of $\beta$: there exists an intermediate
regime ($\beta \sim 1$) where the statistics are dominated by the
non integrability of the dynamics, and are of the G.O.E. type.
In the low temperature limit and in the high temperature limit,
no eigenvalue repulsion was found.
Another important feature of the spectrum of the Glauber matrix
is that for disordered systems, the eigenvalues are close to integers
in the zero temperature limit. This is a consequence of the fact that
the local field never vanishes on any site. This property is not
restricted to disordered systems, since we also observe eigenvalue
clustering at low temperatures for the nearest neighbor Ising
model on the ferromagnetic Cayley tree. Notice that the clustering
property is not related to glassiness since disordered systems
without glassiness may exist (the biased SK model with a sufficient
ferromagnetic bias) and ferromagnets exist which are known
to be glassy at low temperatures and which are not expected to
have this property (regular fractals, percolation clusters,
hyperbolic lattices).

We also studied the separation between the two largest eigenvalues.
This quantity exhibits strong finite size effects. In the case
of the one dimensional Ising model, the frustrated one dimensional model,
and the unbiased SK model,
this quantity decays approximately exponentially with the inverse
temperature. However, in the case of the two dimensional Ising model
and in the case of a biased SK model,
the decay is faster than exponential. This suggests
that the separation between the two largest eigenvalues is sensitive
to the existence of a second order transition. In the thermodynamic limit,
$\Delta_N(\beta)$ is expected to diverge at the transition temperature
in a semi-log plot.

We also studied the density of zero eigenvalues in the SK model.
At zero temperature, this quantity increases with the number of sites.
We found variations that are compatible with an exponential growth.
However, it would be interesting to investigate larger systems.
As far as the highest non-zero eigenvalue probability distribution
is considered, we found that at low temperature, the standard
deviation of this distribution is of the order of the average value
of the distribution, whereas in the high temperature regime,
the standard deviation is small compared to the average value.
This result is compatible with the fact that glassiness is due
to an accumulation of eigenvalues in the vicinity of zero.

In the present study, we only treated the case of small systems.
A generalization to larger systems is under way, in which it
would be especially
interesting to study the separation between the two largest
eigenvalues, as well as the probability of occurence
of zero eigenvalues at zero (and non zero) temperatures.
Another important point is that it would be interesting
to have analytic expressions for $\Delta_N(\beta)$. This question is
also under study. We shall also address in the near future the case
of a Glauber dynamics with an external magnetic field.

I thank B. Dou\c{c}ot for a critical reading of the manuscript,
and I am indebted to J.C. Angl\`es d'Auriac and H. Bruus for
help with group theory. I also thank C. Denniston for reading the
manuscript. I thank the referee for pertinent criticism.

\newpage
\renewcommand\textfraction{0}
\renewcommand
\floatpagefraction{0}
\noindent {\bf Figure captions}

\begin{figure}[h]
\caption{}
\label{Fig1}
Evolution of the eigenvalues of the Glauber matrix as a function
of the inverse temperature for: (a) an 8 site Ising chain in the
reflection antiperiodic sector and the Z2 antiperiodic sector.
No avoided crossings are present, as expected for an integrable system;
(b) the 3x4 Ising model on a square lattice,
in the representation number 3 and the antisymmetric $Z_2$ sector.
The evolution of 32 eigenvalues is plotted on this figure;
(c) the frustrated one dimensional model with nearest neighbor
interactions, for 8 sites in the antisymmetric reflection sector,
and the antisymmetric $Z_2$ sector.
\end{figure}

\begin{figure}[h]
\caption{}
\label{Fig2}
Evolution of the parameter $\Delta_N(\beta)$ as a function of the
inverse temperature $\beta$, for different sizes (open Ising chain).
\end{figure}

\begin{figure}[h]
\caption{}
\label{Fig3}
Eigenvalue spacing statistics of the one dimensional Ising model
with nearest neighbors couplings. The inverse temperature is
$\beta=1$. The statistics are not universal.
\end{figure}

\begin{figure}[h]
\caption{}
\label{Fig5}
Evolution of $\Delta_N(\beta)$ as a function of the inverse temperature
$\beta$ for the 3x4 lattice ($N=12$) and the 3x3 lattice ($N=9$).
Finite size effects are visible, and the variations of $\Delta_N(\beta)$
are not exponential.
\end{figure}

\begin{figure}[h]
\caption{}
\label{Fig6}
Eigenvalue spacing statistics of the 3x4 square lattice for $\beta=0.01$
and $\beta = 1$.
The Poisson and G.O.E. distribution are plotted in dashed lines.
In the regime where initially degenerate eigenvalues
spread linearly ($\beta=0.01$), $P(0)
\simeq 0.5$ and the statistics are close to the Poisson law for large $s$.
In the eigenvalue repulsion regime ($\beta=1$), the statistics exhibit
eigenvalue repulsion ($P(0)=0$), but the shape of $P(s)$ is distinctly
different from the G.O.E. shape.
\end{figure}

\begin{figure}[h]
\caption{}
\label{Fig7}
Eigenvalue spacing statistics of the 3x4 square lattice for $\beta=10$.
The Poisson distribution is plotted in dashed lines.
\end{figure}

\begin{figure}[h]
\caption{}
\label{Fig8}
The one dimensional Ising model with next-nearest-neighbor coupling,
and its representation as a succession of triangles.
The dashed lines represent how the sites are gathered in the
transfer matrix formalism.
\end{figure}

\begin{figure}[h]
\caption{}
\label{Fig9}
Evolution of $\Delta_N(\beta)$ as a function of $\beta$
for $N=6,7,8,9,10$ in the case of the disordered one dimensional model
of Figure \ref{Fig8}. If $N>N'$, $\Delta_N(\beta)<\Delta_{N'}(\beta)$;
$\Delta_N(\beta)$ decreases exponentially with the inverse temperature
$\beta$.
\end{figure}

\begin{figure}[h]
\caption{}
\label{Fig11}
Eigenvalue spacing statistics of the frustrated one dimensional model
with next nearest neighbor interactions, for an inverse
temperature $\beta=0.01$ and $\beta=1$.
\end{figure}

\begin{figure}[h]
\caption{}
\label{Fig12}
Eigenvalue spacing statistics of the frustrated one dimensional model
with next nearest neighbor interactions, for an inverse
temperature $\beta=3$.
\end{figure}

\begin{figure}[h]
\caption{}
\label{Fig13}
Evolution of the eigenvalues as a function of the inverse temperature $\beta$
(a) in the $Z_2$ antisymmetric sector of a 7 site SK model,
and for a given disorder configuration;
(b) in the $Z_2$ antisymmetric sector of the 7 site
SK model with a ferromagnetic bias,
and for a given disorder configuration. The
ferromagnetic bias is $\delta=1.5$.
\end{figure}

\begin{figure}[h]
\caption{}
\label{Fig14}
Evolution of $\overline{\Delta_N(\beta)}$ as a function of $\beta$
for $N=9$ in the case of the SK model, for different values of the
ferromagnetic bias $\delta$. $\overline{\Delta_N(\beta)}$ has been
averaged over $200$ realizations of the disorder. We plotted
 $\overline{\Delta_N(\beta)}$ for $\delta=0,1,1.5,2,3$.
$\overline{\Delta_N(\beta)}$ decreases if the ferromagnetic bias $\delta$
increases, which is consistent with the fact that the transition temperature
increases as a function of $\delta$.
\end{figure}

\begin{figure}[h]
\caption{}
\label{Fig15}
Eigenvalue spacing statistics of the SK model for an inverse temperature
$\beta=.01$ and $\beta=0.1$.
The G.O.E. shape and the Poisson shape are plotted
in dotted lines.
\end{figure}

\begin{figure}[h]
\caption{}
\label{Fig16}
Eigenvalue spacing statistics of the SK model for an inverse temperature
$\beta=1$ and $\beta=10$.
The G.O.E. shape and the Poisson shape are plotted
in dotted lines.
\end{figure}

\begin{figure}[h]
\caption{}
\label{Fig17}
Probability to find a zero eigenvalue at zero temperature, as a function
of the number of sites, for $\delta=-2$,$\delta=0$ and $\delta=2$.
This is a semi-log plot.
\end{figure}

\begin{figure}[h]
\caption{}
\label{Fig18}
Distribution of the largest non-zero eigenvalue $-\Delta_N(\beta)$
for a 9 site SK model, and for $\beta=0.2$ and $\beta=1.7$.
At low temperatures, the standard deviation is of the same order of
magnitude as the mean value, whereas this is not the cas at high temperatures.
\end{figure}

\begin{figure}[h]
\caption{}
\label{Fig19}
Eigenvalue spacing statistics of the biased SK model for an inverse temperature
$\beta=.01$ and $\beta=0.1$. The bias is $\delta=1.5$.
The G.O.E. shape and the Poisson shape are plotted
in dotted lines.
\end{figure}

\begin{figure}[h]
\caption{}
\label{Fig20}
Eigenvalue spacing statistics of the biased SK model for an inverse temperature
$\beta=1$. The bias is $\delta=1.5$.
The G.O.E. shape and the Poisson shape are plotted
in dotted lines.
\end{figure}

\clearpage

\begin{figure}[h]
\caption{}
\label{Fig22}
Eigenvalues of the Glauber matrix on the Cayley tree with 2 generations
and coordination 3 (10 sites). The inverse temperature is $\beta=2$.
\end{figure}


\begin{thebibliography}{99}
\bibitem{ref3} We used the program GAP to determine the classes and
to construct the table of characters.
\end{thebibliography}
\end{document}